\documentclass[preprint,eqsecnum,preprintnumbers,nofootinbib,byrevtex,prd,aps,showpacs,showkeys,groupedaddress,floatfix]{revtex4}
\usepackage{graphics}
\usepackage{graphicx}
\usepackage{epsfig}
\usepackage{amssymb}
\usepackage{amsmath}
\usepackage{enumitem}

\newcommand\nn{\nonumber}
\newcommand\ba{\begin{eqnarray}}
\newcommand\ea{\end{eqnarray}}

\begin{document}

\title{Direct $\rho_L$ meson production in proton-proton and proton-antiproton collisions}

\author{C.~Aydin$^{1}$~\footnote{E-mail: coskun@ktu.edu.tr}}
\author{O.~Uzun$^{1}$~\footnote{E-mail:oguzhan\_deu@hotmail.com}}
\author{A.~I.~Ahmadov$^{2,3}$~\footnote{E-mail: ahmadovazar@yahoo.com}}

\affiliation {$^{1}$ Department of Physics, Karadeniz Technical
University, 61080, Trabzon, Turkey\\
$^{2}$ Department of Theoretical Physics, Baku State University, Z.
Khalilov st. 23, AZ-1148, Baku, Azerbaijan\\$^{3}$Institute  for
Physical Problems, Baku State University,\\ Z. Khalilov st. 23,
AZ-1148, Baku, Azerbaijan}

\begin{abstract}

We have investigated the contribution of the Higher Twist (HT)
Feynman diagrams to the large-$p_T$  inclusive $\rho_L$ meson
production cross section in proton-proton and proton-antiproton
collisions and we discuss the phenomenological consequences of
possible HT contributions to cross sections. To extract
HT subprocesses from Leading Twist (LT) background, we use
various  $\rho_L$ meson distribution amplitudes(DAs). In the numerical
calculations, the dependencies of the HT contribution on the
transverse momentum and the rapidity are discussed with special
emphasis on DAs. Analysis of our results shows that HT contributions
decrease more rapidly than  LT contributions with increasing $p_T$. The
preceding results demonstrate that HT contributions must
be considered especially in the region of low $p_T$ and the HT contribution
to the cross-section depends on the choice of the meson distribution amplitudes.

\end{abstract}


 \maketitle

\section{INTRODUCTION}
Quantum chromodynamics (QCD) is the basic field theory of strong
interactions used for describing hadronic processes. High- and low-
energy dynamics of a hadronic process at large transverse momentum
$(p_T)$ can be analyzed within the framework of the factorization
theorem \cite{Collins}. According to this theorem, high $p_T$
hadronic processes can be separated into two parts each appearing at
different energy scales. While the hard part that takes place between
the interacting partons can be calculated by perturbative quantum
chromodynamics (pQCD), the soft part that refers to the bound state
dynamics of external hadrons is of nonperturbative nature. It has
been observed that the high $p_T$  hadronic processes such as the production of
hadrons, prompt photons, and hadron jets are not described by the
simple dependence predicted by the original parton model \cite{BBK,Berger}.

The idea of direct hadron production was first considered in
1970s to explain the large fixed scalling exponents reported at ISR
and fixed target FNAL energies \cite{Brodsky1}. Hadron production
studies at large $p_T$ provides a valuable testing ground for the
perturbation regime of QCD and also information about both the
parton distribution functions (PDFs) of hadrons and the parton to
hadron fragmentation functions (FFs). Proton-proton collisions are
known to be the most elementary interactions and form the very basic
of our knowledge about the nature of high-energy particle collisions. For these reasons inclusive
direct longitudinally polarized $\rho_L$ meson production in $pp$
proton-proton and $p\bar p$ proton-antiproton collision processes
has an important place in phenemologic research. Origin of the
contributions to cross section to this processes can be separated
into two parts

\begin{itemize}
\item  Higher twist (HT) subprocesses $q_1+\bar{q}_2\to \rho_{L}^{+}(\rho_{L}^{-})+\gamma$  and,
\item  Leading twist (LT) subprocesses as the background of the HT subprocesses
such as $q+\bar{q}\to\gamma+g$, where the gluon is fragmented to a
$\rho_L$ meson $g\to \rho_{L}^{+}(\rho_{L}^{-})$, and $q+g\to
\gamma+q$, where the quark is fragmented to a $\rho_L$ meson $q\to
\rho_{L}^{+}(\rho_{L}^{-})$, $\bar{q}+g\to\gamma+\bar{q}$, where the
antiquark is fragmented to a $\rho_L$ meson
$\bar{q}\to\rho_{L}^{+}(\rho_{L}^{-})$ and so on.
\end{itemize}
LT is the standard processes of the pQCD, where
hadrons are produced via fragmentation processes indirectly from the partons with the fractional momentum z. In contrast, HT
processes are often understood as the direct hadron production, where
the hadron is produced directly in the hard subprocess without
fragmentation \cite{Owens}. A natural explanation for the large
exponents observed in the hadron channel is the presence of
important HT contributions from processes in which the detected
hadron is produced directly in the hard subprocess due to the hadron distribution amplitude \cite{Arleo}.  In a
general QCD analysis of inclusive hadroproduction, all contributing leading and HT hard subprocesses should be considered
\cite{Brodsky2}. In order to calculate LT contributions, parton interaction cross-sections and relevant distribution
and FFs at the appropriate scales should be known. PDFs and FFs represent
intrinsic constituents of the proton and the hadronization
mechanism, respectively. These functions cannot be calculated using
perturbation theory, therefore, they can be obtained from data for
various types of hard-scattering processes. Due to the emergence
of the hadron in the final state directly at the hard-scattering
subprocess, calculation of HT contributions would require
nonperturbative, processes-independent distribution amplitudes
(DAs), instead of FFs.

DAs describing the distribution of partons inside the hadron, provides
essential information on the nonperturbative structure of a hadron\cite{Ball1}.
The choice of proper hadronic DA is one of the key components of analysis on
the direct hadron production. During the past few decades, several important nonperturbative
tools have been developed, which allow specific predictions for the
hadronic DAs.

A main difficulty in making precise pQCD predictions is
the uncertainty in determining the renormalization scale $\mu_R$ of
the running coupling $\alpha_s(\mu_{R}^2)$ and also in factorization
scale $\mu_F$. In  the practical calculations, it is difficult to guess a
simple physical scale of order of a typical momentum transfer in the
process. In a common case, this problem for all orders is solved in
Refs. \cite{Brodsky3} and \cite{Mojaza}. In the present calculations for
renormalization and  factorization scales  we used momentum squared
carried by the hard gluon, which is obtained directly from Fig1.

Light vector and pseudoscalar meson production provides a
reference for high-energy heavy-ion collisions. In fact, key
information on the hot and dense state of strongly interacting
matter produced in these collisions can be extracted measuring light
meson \cite{ALLICE}. Precision experimental studies of $\rho_L$ meson production in
proton-antiproton and proton-proton collisions at low energies are
proposed in the experiment named  PANDA \cite{PANDA}. The PANDA
scientific program uses low-energy ranging between 1.5-15 GeV for
interactions between protons and antiprotons, where this energy lies
around the pion and $\rho$ meson production threshold. This program
includes several measurements and it addresses fundamental questions
of QCD by obtaining the detailed analyses of all possible mechanisms
of meson pair production \cite{Poslavsky}.

Many research papers can be found in the scientific literature on
the contribution of the HT effects to the cross-section. Inclusive
gluon production in pion-proton collisions \cite{Ahmadov1}, direct
photon production in a pion-proton collision \cite{Ahmadov2}, pion
production in proton-proton and photon-photon collisions
\cite{FK1,FK2,Du} are some examples of these studies.  Direct
longitudinally polarized $\rho_{L}$  meson production in the  $pp$
proton-proton and $p\bar p$ proton-antiproton collision can be used
to tune particle production models and they are the key probes of the
hot and dense state of strongly interacting matter produced in
heavy-ion collisions. Also, longitudinally polarized $\rho_{L}$
meson production in the $p\bar p$ collision is a very important task of
the FAIR experiment.

In this study, we examine the contribution of the HT effects to
inclusive $\rho_{L}^{\pm}$ production at proton-antiproton and
proton-proton collisions by using different $\rho_{L}$ meson DAs
which can be helpful for an explanation of the PENIUX and PANDA
experiments. We have also given theoretical predictions of the
inclusive $\rho_{L}^{\pm}$ meson production in $p\bar{p}$ and $pp$
collisions by accounting for the leading order diagrams in partonic
cross-sections.

We show that the HT terms contribute substantially to the
inclusive meson cross-section at moderate transverse momenta. In
addition, we demonstrate that HT reactions necessarily
dominate in the kinematic limit where the transverse momentum
approaches the phase-space boundary. Another important aspect of
this study is the choice of the $\rho_L$ meson DAs. In this respect,
the HT Feynman diagrams have been computed by using various
$\rho_{L}$ meson DAs obtained by using ligth front quarks model,QCD sum rules and lattice QCD.

The rest of the paper is organized as follows: In Sec.\ref{lt}, formulas
for LT cross-sections for $\rho_{L}$ meson production are provided. In Sec.~\ref{ht},  a
brief information for the calculation of the HT
contribution to cross-section and the formulas for the HT cross-
section of the process $pp\to\rho_{L}\gamma X$  and $p\bar
{p}\to\rho_{L}\gamma X$ are given.
Numerical results for the cross-section and the discussion of the
dependence on the cross-section on the $\rho_{L}$ meson DA are
provided in Sec.~\ref{results}. Finally, the concluding remarks are
stated in Sec.~\ref{conc}.

\section{CONTRIBUTION OF THE  LEADING TWIST DIAGRAMS}\label{lt}
The LT subprocesses for the $\rho_{L}$  meson production
we take two subprocesses: (1) quark-antiquark annihilation $q\bar{q}
\to g\gamma$, in which the $\rho_{L}$ meson indirectly emitted from
the gluon, $g \to \rho_{L}^{+}(\rho_{L}^{-})$ and (2) quark-gluon
fusion, $qg \to q\gamma $, with subsequent fragmentation of final
quark into a meson, $q \to \rho_{L}^{+}(\rho_{L}^{-})$. The
Mandelstam invariant variables for subprocesses $q_1+\bar{q}_{2} \to
\rho_{L}^{+}(\rho_{L}^{-})+\gamma$ presented in Fig.1 are defined as
\begin{equation}
\hat s=-(p_1+p_2)^2,\quad \hat t=-(p_1-p_{\rho})^2,\quad \hat
u=-(p_1-p_{\gamma})^2.
\end{equation}

In our calculations, the quarks masses are neglected. The corresponding cross-
sections is obtained as
\begin{equation}
\frac{d\sigma}{d\hat t}(q\bar{q} \to g\gamma)=\frac{8}{9}\pi\alpha_E
\alpha_s(Q^2)\frac{e_{q}^2}{{\hat s}^2}\left(\frac{\hat t}{\hat
u}+\frac{\hat u}{\hat t}\right),
\end{equation}
\begin{equation}
\frac{d\sigma}{d\hat t}(qg \to q\gamma)=-\frac{\pi{e_{q}^2}\alpha_E
\alpha_s(Q^2)}{{3\hat s}^2}\left(\frac{\hat s}{\hat t}+\frac{\hat
t}{\hat s}\right),
\end{equation}
For the LT contribution, we found
%
\ba
\Sigma_{M}^{LT}\equiv E\frac{d\sigma}{d^3p}=\sum_{q}\int_{0}^{1}
dx_1 dx_2dz \biggl(G_{{q_{1}}/{h_{1}}}(x_{1},Q_{1}^2)
G_{{q_{2}}/{h_{2}}}(x_{2},Q_{2}^2)D_{g}^{\rho}(z)\frac{\hat s}{\pi
z^2}\frac{d\sigma}{d\hat t}(q\bar{q}\to g\gamma)+ \nn \\
G_{{q_{1}}/{h_{1}}}(x_{1},Q_{1}^2)
G_{{g}/{h_{2}}}(x_{2},Q_{2}^2)D_{q}^{\rho}(z)\frac{\hat s}{\pi
z^2}\frac{d\sigma}{d\hat t}(qg \to q\gamma)\biggr) \delta(\hat
s+\hat t+\hat u), ~~
\ea
where
\begin{equation}
\hat s=x_{1}x_{2}s,\,\,\hat t=\frac{x_{1}t}{z},\,\,
\hat u=\frac{x_{2}u}{z},\,\, z=-\frac{x_{1}t+x_{2}u}{x_{1}x_{2}s}.
\end{equation}
$G_{q_{1,2}/h_{1,2}}(x,Q^2)$ and $G_{g/ h}(x,Q^2)$ are the quark(antiquark) and gluon
distribution function inside a proton and antiproton, respectively.
$D_{g}^{\rho}(z)=D_{g}^{\rho_{L}^{+}}(z)=D_{g}^{\rho_{L}^{-}}(z)$
and $D_{q}^{\rho_{L}}(z)$ are the gluon and the
quark  fragmentation function. The $\delta$ function may be expressed in terms of the
parton kinematic variables, and the $z$ integration may then be
done. The final form for the cross section is obtained as
\ba
\Sigma_{M}^{LT} \equiv
E\frac{d\sigma}{d^3p}=\sum_{q}\int_{x_{1min}}^{x_{1max}} dx_1
\int_{x_{2min}}^{x_{2max}} dx_2\int_0^1 dz
\biggl(G_{{q_{1}}/{h_{1}}}(x_{1},Q_{1}^2)
G_{{q_{2}}/{h_{2}}}(x_{2},Q_{2}^2) \cdot \nn \\ 
\cdot D_{g}^{\rho_{L}}(z) \frac{\hat s}{\pi z^2}\frac{d\sigma}{d\hat t}(q\bar{q} \to g\gamma)
+ G_{{q_1}/{h_1}}(x_1,Q_{1}^2) G_{g/{h_2}}(x_{2},Q_{2}^2)D_{q}^{\rho_L}(z) \cdot
\frac{\hat s}{\pi z^2}\frac{d\sigma}{d\hat t}(qg \to q\gamma)\biggr) = \hspace{0.5cm} \nn \\
= \sum_{q}\int_{x_{1min}}^{x_{1max}} dx_1 \int_{x_{2min}}^{x_{2max}}
\frac{dx_2}{-(x_{1}t+x_{2}u)}\biggl(x_{1}G_{{q_{1}}/{h_{1}}}(x_{1},Q_{1}^2)
sx_{2}G_{{q_{2}}/{h_{2}}}(x_{2},Q_{2}^2)\frac{D_{g}^{\rho_{L}}(z)}{\pi} \cdot \nn \\
\cdot \frac{d\sigma}{d\hat t}(q\bar{q} \to g\gamma) +   
x_{1}G_{{q_{1}}/{h_{1}}}(x_{1},Q_{1}^2)
sx_{2}G_{{g}/{h_{2}}}(x_{2},Q_{2}^2)\frac{D_{q}^{\rho_{L}}(z)}{\pi}
\frac{d\sigma}{d\hat t}(qg \to q\gamma)\biggr), ~~~~
\ea
where
$$
x_{1min} =\frac{p_{T}e^{y}}{\sqrt s-p_{T}e^{-y}},\,\,\, x_{2min}=\frac{x_1p_{T}e^{-y}}{x_{1}\sqrt s-p_{T} e^{y}}.
$$
In really in Eq.(2.6) the region of integration we took from  $x_{1min}$ to $x_{1max}$, and $x_{2min}$ to $x_{2max}$
which here $x_1$ is changed in this interval  $x_1\in[0.033\div0.925]$, so  $x_{1min}=0.033$, $x_{1max}=0.925$ and
$x_2$ is changed in this interval  $x_2\in[0.001\div0.857]$, so $x_{2min}=0.001$, $x_{2max}=0.857$. Thus,  here $x_{1max}$ is
not equal to unite, so  $(x_{1max}-1)\neq 0$. Therefore, here $\alpha_{s}$  have not any infrared singularity

\section{HIGHER TWIST  CONTRIBUTIONS TO INCLUSIVE DIRECT\\ LONGITUDINALLY POLARIZED  $\rho_{L}$ MESON PRODUCTION}\label{ht}

The HT Feynman diagrams  are shown in Fig.1. According to factorization theorem, the HT amplitude $M$
can be factored in terms of the elementary hard-scattering amplitude
$(T_H)$ and hadron DA $(\Phi_{\rho_L})$ for
directly produced $\rho_L$ meson. We have
aimed to calculate the $\rho_L$  meson production cross-section and to fix
the differences due to the use of various $\rho_L$ meson DAs.
The amplitude for this subprocess can be found by
means of the Brodsky-Lepage formula~\cite{Lepage2}
\ba M(\hat s,\hat
t)=\int_{0}^{1}{dx_1}\int_{0}^{1}dx_2\delta(1-x_1-x_2)\Phi_{M}(x_1,x_2,Q^2)T_{H}(x_1,x_2;Q^2,\mu_{R}^2,\mu_{F}^2)
\label{BL} \ea
It is well know that, the hard-scattering amplitude $T_{H}(x_1,x_2;Q^2,\mu_{R}^2,\mu_{F}^2)$ in Eq.\eqref{BL} depends
on a process and can be obtained in the framework of pQCD and
represented as a series in the QCD running coupling constant
$\alpha_{s}(Q^2)$. The light-cone momentum fractions $x\equiv x_1$,
$x_2=1-x$ specify the fractional momenta carried by quark and
antiquark in the Fock state.

At the leading order of pQCD calculations, the hard scattering
amplitude $T_{H}(x_1,x_2;Q^2, \mu_{R}^2,\mu_{F}^2)$ does not depend
on the factorization scale $\mu_{F}^2$, but strongly depends on renormalization scale
$\mu_{R}^2$. However, the scales $\mu_{F}^2$ and $\mu_{R}^2$ are
independent of each other. Therefore, it does not
depend on the choice of the renormalization scale or the
factorization scale.

Note that in our calculations for renormalization scale we use the
momentum squared carried by the hard gluon (virtuality of gluon) which can
be seen directly from Fig.1. Natural way, to obtain the expression for the
momentum square is comes from the application of the four-momentum conservation
law to Fig.1. We eventually find the expression  for $Q_{1}^{2}(x)$ as:
$Q_{1}^{2}(x)=-(1-x_1)\hat u$ while $Q_{2}^{2}(x)=-x_1 \hat t$.

Thus, we have shown the reason of the choosing renormalization scale.
The rest of the results we have presented in this paper are based on the above reasoning.
Thus as noted in above, the choosing  renormalization scale (virtuality
 of hard gluon) is not conjectural.

In calculations of $T_H$, we have neglected external lines and also
rho meson mass. Furthermore, we considered the fact that initial quarks
combine to form color neutral particles, which are restricted to be in the
spin state of the final rho meson.

The cross-section for the HT subprocess  is given by the
expression
\begin{equation}
\frac{d\sigma}{d\hat t}(\hat s,\hat t,\hat u)=\frac
{8\pi^2\alpha_{E} C_F}{27}\frac{\left[G(\hat t,\hat
u)\right]^2}{{\hat s}^3}\left[\frac{1}{{\hat u}^2}+\frac{1}{{\hat
t}^2}\right],
\end{equation}
where
\begin{equation}
G(\hat t,\hat u)=e_1\hat
t\int_{x_{1min}}^{x_{1max}}dx_1\left[\frac{\alpha_{s}(Q_1^2)\Phi_{\rho_{L}}(x_1,Q_1^2)}{1-x_1}\right]+e_2\hat
u\int_{x_{1min}}^{x_{1max}}dx_1\left[\frac{\alpha_{s}(Q_2^2)\Phi_{\rho_{L}}(x_1,Q_2^2)}{1-x_1}\right],
\end{equation}
where $e_1(e_2)$ is the charge of $q_1(\bar{q}_2)$
and $C_F=\frac{4}{3}$.

First, it seems that even the integral (3.3) has a singularity, since the $\rho$ meson function ($\Phi_\rho$)
is the order of $(1-x)$ (see Eqs.(3.9 - 3.13)) it does not have any divergency problem.
Another side in Eq.(3.3), in the region of integration, we adopt from  $x_{1min}$ to $x_{1max}$, in
which here $x_1$  is changed in this interval  $x_1\in[0.033\div0.925]$, also here $x_{1max}$ is
not equal to unite, so  $(x_{1max}-1)\neq 0$.
According of the values $x_{1,2}$, the minimal value of $Q_{1,2}^2$ accepts the value of $(120.668, 4.237)$.
Therefore, here,  $\alpha_s$ has not got any infrared singularity.
The HT contribution to the large-$p_{T}$ longitudinally polarized $\rho_{L}$ meson production
cross-section in the process $pp\to\rho_{L}^{\pm}\gamma X$ \cite{Owens} is
\ba
\Sigma_{M}^{HT}\equiv E\frac{d\sigma}{d^3p}= \nn \\ 
=\int_{0}^{1}\int_{0}^{1}
dx_1 dx_2 G_{{q_{1}}/{h_{1}}}(x_{1},Q_{1}^2)
G_{{q_{2}}/{h_{2}}}(x_{2},Q_{2}^2)\frac{\hat s}{\pi} \frac{d\sigma}{d\hat
t}(q\overline{q}\to \rho_{L}\gamma)\delta(\hat s+\hat t+\hat u)= 
 {1 \over \pi} \frac{d\sigma}{dydp_{T}^2}. ~~
\ea
The Mandelstam variables  $t$ and $u$ also can be expressed in terms
of the process center-of-mass energy, transverse momentum of
$\rho_L$ meson, and the rapidity by using the following expressions:
$$\hat t=x_1t,$$
\begin{equation}
\hat u=x_2u,
\end{equation}
$$t= -m_T \sqrt{s} e^{-y}=-p_T \sqrt{s}e^{-y},$$
$$u= -m_T \sqrt {s} e^y=-p_T \sqrt{s}e^{y}.$$

Using the relation in Eq.(3.5) and the fact that for the massles
two-body scattering $\hat s+\hat t +\hat u =0$, we obtain
$$
x_1=-\frac{x_{2}u}{x_{2}s+t}=\frac{x_{2}p_{T}\sqrt s
e^{y}}{x_{2}s-p_{T}\sqrt s e^{-y}},
$$
$$
x_2=-\frac{x_{1}t}{x_{1}s+u}=\frac{x_{1}p_{T}\sqrt s
e^{-y}}{x_{1}s-p_{T}\sqrt s e^{y}},
$$

where $m_T$ is the transverse mass of $\rho_L$  meson, which is
given by
$$m_T^2=m^2+p_T^2.$$

For a full discussion and for the possible extraction of the HT
contributions to cross-section, we calculated difference of the
cross-sections of the longitudinally polarized $\rho_L$ meson
production in proton-antiproton and proton-proton collisions. It
should be noted that a similar approach is widely applied in the
extraction of the contribution of the spin effects to the one and
double spin asymmetries:
\begin{align}
&\Delta=\Sigma_{p\bar p}-\Sigma_{pp}
\end{align}
where,
\begin{align}
&\Sigma_{p\bar p}=E_{\rho_{L}^+}\frac{d^3\sigma}{d^3p} \left(p\bar
p\rightarrow\rho_{L}^{+} +\gamma +X \right)
+E_{\rho_{L}^-}\frac{d^3\sigma}{d^3p}\left(p\bar
p\rightarrow\rho_{L}^{-} +\gamma +X \right)
\end{align}

and

\begin{align}
&\Sigma_{pp}=E_{\rho_{L}^+}\frac{d^3\sigma}{d^3p}\left(p
p\rightarrow\rho_{L}^+ +\gamma +X \right)
+E_{\rho_{L}^-}\frac{d^3\sigma}{d^3p}\left(pp\rightarrow\rho_{L}^-
+\gamma +X \right)\nonumber
\end{align}

In this study, we utilized the following $\rho_{L}$ meson DAs:
asymptotic DAs  $\Phi_{\rho_L}^{asy}(x)$, derived in pQCD
evaluation, Ball-Braun $\Phi_{\rho_L}^{BB}(x,Q^2)$ DA obtained by
using QCD sum rules ~\cite{Ball},  $\Phi{\rho_L}^{PMS}(x,Q^2)$ is
obtained by Pimikov \emph{et al}. \cite{Pimikov} by using
QCD sum rules, $\Phi_{\rho_L}^{Linear}(x,Q^2)$, and
$\Phi_{\rho_L}^{HO}(x,Q^2)$ DAs obtained by  Choi and Ji
\cite{Choi} from  linear and harmonic oscillator potential model,
respectively.

The principle definition of DA of the $\rho$ meson have the form ~\cite{Ball}:
\ba
<0\mid\bar{u}(0)\gamma_{\mu}d(x)\mid\rho^{+}(p,\lambda)>=p_{\mu}\frac{(e^{(\lambda)}x)}{(px)}f_{\rho}m_{\rho}\int_{0}^{1}
du e^{-iupx}\Phi_{\rho_L}(u,\mu) \nn  \\
(e_{\mu}^{(\lambda)}-p_{\mu}\frac{(e^{(\lambda)}x)}{(px)})f_{\rho}m_{\rho}\int_{0}^{1}
du e^{-iupx}g_{\bot}^{(\upsilon)}(u,\mu)
\label{aDef} \ea

Should be noted that for construction DA for the light mesons $\Phi_{\rho}(x, \mu_{0}^2)$ is taken massless and normalized to unit.
They are defined by the following expressions:
\ba \Phi_{\rho_L}^{asy}(x)=6x(1-x)\label{asy} \ea
\ba
\Phi_{\rho_L}^{BB}(x,\mu_{0}^2)=\Phi_{\rho_L}^{asy}(x)\left[C_{0}^{3/2}(2x-1)+0.18\cdot C_{2}^{3/2}(2x-1)\right], \label{rhob}\ea
\ba
\Phi_{\rho_{L}}^{PMS}(x,\mu_{0}^2)=\Phi_{\rho_{L}}^{asy}(x)\biggl[C_{0}^{3/2}(2x-1)+0.047\cdot C_{2}^{3/2}(2x-1)-  \nn  \\
-0.057\cdot C_{4}^{3/2}(2x-1)\biggr], \label{PMS} \ea
\ba
\Phi_{\rho_{L}}^{Linear}(x,\mu_0^2)=\Phi_{\rho_L}^{asy}(x)\biggl[C_{0}^{3/2}(2x-1)+0.02\cdot C_{2}^{3/2}(2x-1)
-0.01\cdot C_{4}^{3/2}(2x-1)-  \nn \\
 0.02 \cdot C_{6}^{3/2}(2x-1) \biggr], ~~~~~~ \label{Linear1} \ea
\ba
\Phi_{\rho_{L}}^{HO}(x,\mu_0^2)=\Phi_{\rho_L}^{asy}(x)\biggl[C_{0}^{3/2}(2x-1)-0.02\cdot C_{2}^{3/2}(2x-1)
-0.03 \cdot C_{4}^{3/2}(2x-1)- \nn \\
0.02\cdot C_{6}^{3/2}(2x-1)\biggr], ~~~~~~\label{HO} \ea
where $C_{n}^{3/2}(2x-1)$ are Gegenbauer  polynomials. In the numerical calculations for normalization scale $\mu_{0}^2$ we adopt $\mu_{0}^2=1 GeV^2$.
Also, non-trivial Gegenbauer moments  $a_n$ have been taken at the scale $\mu_{0}^2=1 GeV^2$.

The DAs of mesons, specially of $\pi$ and $\rho$  meson also are developed in Refs.
\cite{Bakulev1,Bakulev2,Bakulev3,Mikhailov5,Dorokhov6,Pimikov1} by
the Dubna group.
Some properties  of Gegenbauer  polynomials given in App. A

\section{NUMERICAL RESULTS AND DISCUSSION}\label{results}

In this section, we discuss in detail the numerical predictions of
the LT and HT cross-sections of the direct $\rho_{L}$ meson
production processes $pp\to\rho_{L}\gamma X$  and
$p\bar{p}\to\rho_{L}\gamma X$ at the $62.4$ GeV energies taking into
account the full leading-order contributions from quark-antiquark
annihilation process. For the HT subprocess, we take
$q_1+\bar{q}_{2} \to (q_1\bar{q}_2)+\gamma$ and for the dominant
LT subprocess for the $\rho_{L}$ meson production, we
take the quark-antiquark annihilation $q\bar{q} \to g\gamma$, in
which the $\rho_{L}^{\pm}$ meson is indirectly emitted from the
gluon  and quark-gluon fusion, $qg \to q\gamma $, with subsequent
fragmentation of final quark into a meson, $q \to \rho_{L}^{\pm}$.
We denote the HT cross-section by $\Sigma_{\rho}^{HT}$, the LT
cross-section by $\Sigma_{\rho}^{LT}$. For the quark and gluon
distribution functions inside the proton and antiproton, the
MSTW2008 PDFs \cite{watt} and the quark and gluon fragmentation
functions \cite{saveetha}  are used. Also, the following
abbreviations are defined for DAs: asy is $\Phi_{\rho_L}^{asy}(x)$,
BB is $\Phi_{\rho_L}^{BB}(x,Q^2)$, PMS is
$\Phi_{\rho_{L}}^{PMS}(x,Q^2)$, Linear is
$\Phi_{\rho_{L}}^{Linear}(x,Q^2)$, and HO is
$\Phi_{\rho_L}^{HO}(x,Q^2)$. The results are given for $\sqrt s$ =
62.4 GeV  on the transverse momentum $p_T$ ranging from 2
GeV/\emph{c} to 30 GeV/\emph{c} which are also valid for the PHENIX
experiment. Obtained results are visualized through Figs.
\ref{fig:fig2}-\ref{fig:fig11}.

First of all, it is very interesting to compare the  HT
cross-sections for all DAs of the $\rho$ meson. In Figs.
\ref{fig:fig2}- \ref{fig:fig8}, we show the HT cross-section
$\Sigma_{\rho_{L}^+}^{HT}$ and $\Sigma_{\rho_{L}^-}^{HT}$ of the
process   $p\bar{p}\to\rho_{L}^{\pm} \gamma X$,  the difference of the
cross-sections $\Delta=\Sigma_{p\bar p}-\Sigma_{pp}$ and the ratio $\Sigma_{\rho}^{HT}/\Sigma_{\rho}^{LT}$, and
$\Delta/\Sigma_{\rho}^{LT}$ as a function of the meson transverse
momentum $p_{T}$ for the  DAs which is presented by Eqs. (\ref{asy})
- (\ref{HO}) and for $y=0$.

As is seen from    Figs. \ref{fig:fig2}- \ref{fig:fig5} the HT  $\Sigma_{\rho_{L}^+}^{HT}$ and $\Sigma_{\rho_{L}^-}^{HT`}$
cross-sections are monotonically decreasing with an increase in the
transverse momentum of the $\rho_L$ meson. It is worth to mention
that at the c.m. energy $\sqrt s$=62.4\,\, GeV the maximum value of
the cross-section of the process $p\bar{p}\to\rho_{L}^{+}\gamma X$ for
the $\Phi_{\rho_L}^{BB}(x,Q^2)$ decreases from the interval
$1.7\times10^{-4}$ mb/GeV$^{2}$ to $5.352\times10^{-20}$
mb/GeV$^{2}$, but the maximum value of
the cross-section of the process $p p\to\rho_{L}^+\gamma X$ for the
$\Phi_{\rho_L}^{BB}(x,Q^2)$ decreases from the interval
$1.992\times10^{-4}$ mb/GeV$^{2}$ to $ 2.634\times10^{-24}$
mb/GeV$^{2}$.
The magnitude for the difference of the cross-sections $\Delta$  for
$\Phi_{\rho_L}^{BB}(x,Q^2)$ decreases from the interval $
9.757\times10^{-6}$ mb/GeV$^{2}$ to $5.566\times10^{-20}$
mb/GeV$^{2}$. It can be seen that the difference  of cross-sections $\Delta$ is slowly decreasing with an
increase in the transverse momentum of the $\rho_{L}$ meson.

Through Figs. \ref{fig:fig7} and  \ref{fig:fig8}, we have displayed the ratio
$\Sigma_{\rho}^{HT}/\Sigma_{\rho}^{LT}$, and
$\Delta/\Sigma_{\rho}^{LT}$ as a function of the meson transverse
momentum $p_{T}$ for the  DAs which is presented by Eqs. (\ref{asy})
- (\ref{HO}) and for $y=0$. The ratio for all DAs has a minimum at
approximately $p_T=15$GeV/\emph{c} and after this, the value increases
monotonically with an increase in the transverse momentum of the
$\rho_{L}$ meson. Ratio of
$\Sigma_{p\bar{p}}^{HT}/\Sigma_{\rho}^{LT}$ for
$\Phi_{\rho_L}^{BB}(x,Q^2)$ DA is changed the interval between
$1.40839\times10^{4}$  to $1.2903\times10^{6}$, for $\Phi_{\rho_L}^{Linear}(x,Q^2)$ DA is
changed in the interval between $7.8656\times10^{3}$  to $8.2459\times10^{5}$. However, for
ratio $\Delta/\Sigma_{\rho}^{LT}$ for $\Phi_{\rho_L}^{BB}(x,Q^2)$ DA
is changed in the interval between $3.6494\times10^{2}$  to $1.2903\times10^{6}$, for
$\Phi_{\rho_L}^{Linear}(x,Q^2)$, DA is changed in the interval between
$1.937\times10^{2}$  to $8.245\times10^{5}$. As is seen from Fig.\ref{fig:fig7} and 8,  the ratio of
HT to LT cross sections for the dependence on the  $p_T$ has
a similar behavior.

Finally  in Figs. \ref{fig:fig9}-\ref{fig:fig11}, we have presented the HT cross-sections
$\Sigma_{p\bar{p}}^{HT}$, $\Sigma_{pp}^{HT}$ and difference
$\Delta=\Sigma_{p\bar p}-\Sigma_{pp}$ as a function of the rapidity
$y$ of the $\rho_{L}$ meson at c.m.energy $62.4\,\,GeV$ and
$p_T$=4.9\,\,GeV/$c$. At  62.4\,\,GeV and $p_T$=4.9\,\,GeV/$c$, the
meson rapidity lies in the region $-2.52\leq y\leq2.52$. As is seen
from figures, the cross-sections and difference of cross-sections for
all DAs increase with an increase of the $y$ rapidity of the meson
and have a maximum approximately at the point $y=1.5$. After
reaching the maximum, it is decreasing rapidly with an increase in
the transverse momentum of the $\rho_L$ meson. As it is seen from
Fig. \ref{fig:fig10} in the region ($-2.52\leq y\leq 0.96$), the
ratio for all DAs increase and it has a maximum
approximately at the point $y=0.96$. Similarly, the ratio decreases
with an increase in the $y$ rapidity of the $\rho_L$ meson. It can
also be seen that the ratios are sensitive to the choice of the
meson DAs. In Fig.\ref{fig:fig11} we present difference
$\Delta=\Sigma_{p\bar p}-\Sigma_{pp}$ as a function of the rapidity
$y$ of the $\rho_{L}$ meson.

For the HT contribution, it is important to analyze its relative magnitude of
contribution compared to the LT contribution, since only
LT diagrams are commonly considered in usual studies of
the hadron-hadron collision. As seen from the figures, HT
contributions are comparable with the cross-section of LT in the
region $p_T<1$0 GeV/$c$. Moreover, in the region $p_T<5$ GeV/$c$ HT
contributions become more significant.

Thus, it can be concluded that HT cross-section  of the $\rho_{L}$ meson
production in the proton-antiproton and proton-proton  collisions appears in the range
and should be observable at the PHENIX and PANDA experiments.

\section{CONCLUSION}\label{conc}

 We have calculated and analyzed the cross-section of
the direct inclusive longitudinally polarized $\rho_{L}$ meson
production via HT mechanism in the proton-proton and
proton-antiproton collisions. We discussed the phenemenological consequences of possible HT
contributions to cross-sections and also compared  with LT.
Moreover, in order to extract HT subprocesses from LT background, we adopted different $\rho_L$ meson DAs.
In the numerical calculations, dependencies of the HT
contribution on the transverse momentum and the rapidity  of the
$\rho_L$ meson are discussed with special emphasis on
DAs. Analysis of our results show that with
increasing $p_T$, the HT contributions decrease more
rapidly then LT contributions.
The preceding results demonstrate that HT contributions must be considered
especially in the region of low $p_T$. It is shown that the HT contribution
to the cross-section depends on the choice of the
meson distributions amplitudes.  Thus, it is crucial to note that
the contribution of HT effects to cross-section the dependence on the DAs is weak.
Despite that, the HT cross-section obtained in our
study should be observable at hadron collider. Also, the feature of
HT effects  may help theoretical interpretations of the future
experimental data for the direct inclusive vector meson production
cross-section in the proton-proton and proton-antiproton collisions.

It is important to note that the difference of the cross-section
$\Delta=\Sigma_{p\bar p}-\Sigma_{pp}$ is not equal to zero. This
enables to calculate and extract the extension of the HT
to cross-section. It is also one of the most critical points for the
experiment.

It can be noted that the HT processes for large-$p_T$ meson
production have a key enabling contribution, where the $\rho_L$
meson are generated directly in the hard-scattering subprocess,
rather than by gluon and quark fragmentation. Inclusive $\rho_L$
meson production provides an essential test case where HT
contributions dominate those of LT in the certain kinematic regions.
The HT contributions can be utilized to interpret theoretically the
future experimental data for the charged $\rho_L$  meson production in
$pp$ and $p\bar p$ collisions. The results of this study can be
useful to provide a simple test of the short distance structure of
QCD as well as to determine more precise DAs of the $\rho_L$ meson.

By taking these points into account, it may be argued that the
analysis of the HT effects on the dependence of the
$\rho_{L}$ meson DAs  in longitudinally polarized  $\rho_{L}$ meson
production at proton-proton  and proton-antiproton collisions is
significant in both theoretical and experimental studies.
\\
\\
\appendix
\section{Gegenbauer polynomials and its some properties}

The  $\rho_L$ meson DA is symmetric under replacement $x_1-x_2\leftrightarrow x_2-x_1$.
Thus, $x\equiv x_1$, $x_2=1-x$ and $x_1-x_2=2x-1$.
A recurrence relation for Gegenbauer polynomials is:
\ba
nC_{n}^{(\lambda)}(\xi)=2(n+\lambda-1)\xi
C_{n-1}^{(\lambda)}(\xi)-(n+2\lambda-2)C_{n-2}^{(\lambda)}(\xi).
\label{GP} \ea

Also, the few Gegenbauer polynomials in  expression (\ref{asy}) - (\ref{HO}) are as follows:
\ba
C_{0}^{3/2}(2x-1)=1,\,\,C_{2}^{3/2}(2x-1)=\frac{3}{2}(5(2x-1)^2-1),\nn \\
C_{4}^{3/2}(2x-1)=\frac{15}{8}(21(2x-1)^4-14(2x-1)^2+1),\nn \\
C_{6}^{3/2}(2x-1)=\frac{1}{16}(3003(2x-1)^6-3465(2x-1)^4+945(2x-1)^2-35).
\label{GPN} \ea

The $\rho$ meson  DA can be expanded  over the eigenfunctions of the
one-loop Efremov-Radyushkin-Brodsky-Lepage (ERBL) equation
~\cite{Lepage11,Efremov1,Efremov2}
\ba
\Phi_{\rho}(x,Q^2)=\Phi_{asy}(x)\left[1+\sum_{n=2,4..}^{\infty}a_{n}(Q^2)C_{n}^{3/2}(2x-1)\right].\label{a36}
\ea
The evolution of the DA on the factorization scale $Q^2$ is handled
by the functions $a_n(Q^2)$ as \ba
a_n(Q^2)=a_{n}^{\|}(\mu_{0}^2)\left[\frac{\alpha_{s}(Q^2)}{\alpha_{s}(\mu_{0}^2)}\right]^{\gamma_{n}^{\|}/2\beta_0},\label{a37}
\ea
Here, the strong coupling constant $\alpha_{s}(Q^2)$ at the one-loop
approximation is given as
\begin{equation}
\alpha_{s}(Q^2)=\frac{4\pi}{\beta_0{ln}(\frac{Q^2}{\Lambda^2})}.
\end{equation}
where $\Lambda$ is the QCD scale parameter, $\beta_0$ is the QCD
beta function one-loop coefficients defined as
$$
\beta_0=11-\frac{2}{3}n_f.
$$
In  expression \eqref{a37} $\gamma_n$'s are anomalous dimensions
defined by the expression
\ba
\gamma_{n}^{\|}=\frac{8}{3}\left[1-\frac{2}{(n+1)(n+2)}+4\sum_{j=2}^{n+1}
\frac{1}{j}\right].\label{a38}
\ea

So, from expression \eqref{a38} we obtain
\ba
\frac{\gamma_{2}^{\|}}{2\beta_{0}}=\frac{50}{81},\,\,\,\frac{\gamma_{4}^{\|}}{2\beta_{0}}=\frac{364}{405},\,\frac{\gamma_{6}^{\|}}{2\beta_{0}}=\frac{1027}{105}\,
\label{a40} \ea

The Gegenbauer moments  $a_n$  can be determined by using the
Gegenbauer polynomials orthogonality condition \ba \label{eq:ort}
\int^1_{-1}(1-\xi^2)
C_{n}^{3/2}(\xi)C_{n'}^{3/2}(\xi)d\zeta=\frac{\Gamma(n+3)\delta_{nn'}}{n!(n+3/2)}.
\ea The Gegenbauer moments $a_n$ are very practical in studying the
DAs because they form the shape of the corresponding hadron wave
function. They can be derived from theoretical
models or extracted from the experimental data. Besides, these
moments reveal how much the DAs deviate from the asymptotic one.

\section*{\bf Acknowledgments}
We cordially thank S. V. Mikhailov and A. V. Pimikov for many helpful and insightful discussions.

\newpage
\begin{figure}
    \begin{center}
    \includegraphics[trim=10 5 1 1,clip,width=12cm]{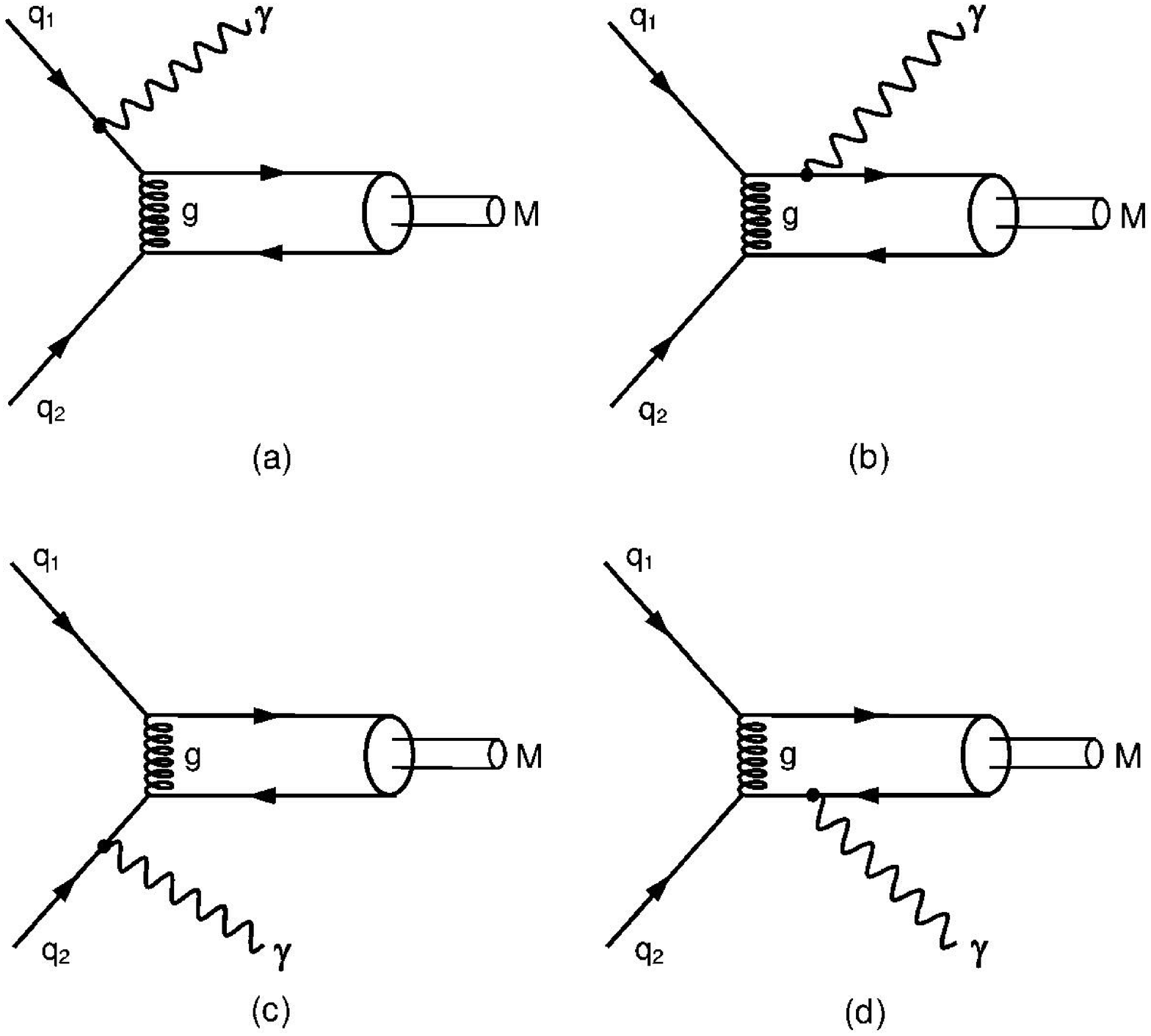}
    \caption{Feynman diagrams for the  HT
subprocess, $q_1 q_2 \to
\rho_{L}^{+}(or\,\,\rho_{L}^{-})\gamma.$.\label{f1}}
\end{center}
\end{figure}

\newpage
\begin{figure}[!htb]
\includegraphics[width=11.8 cm]{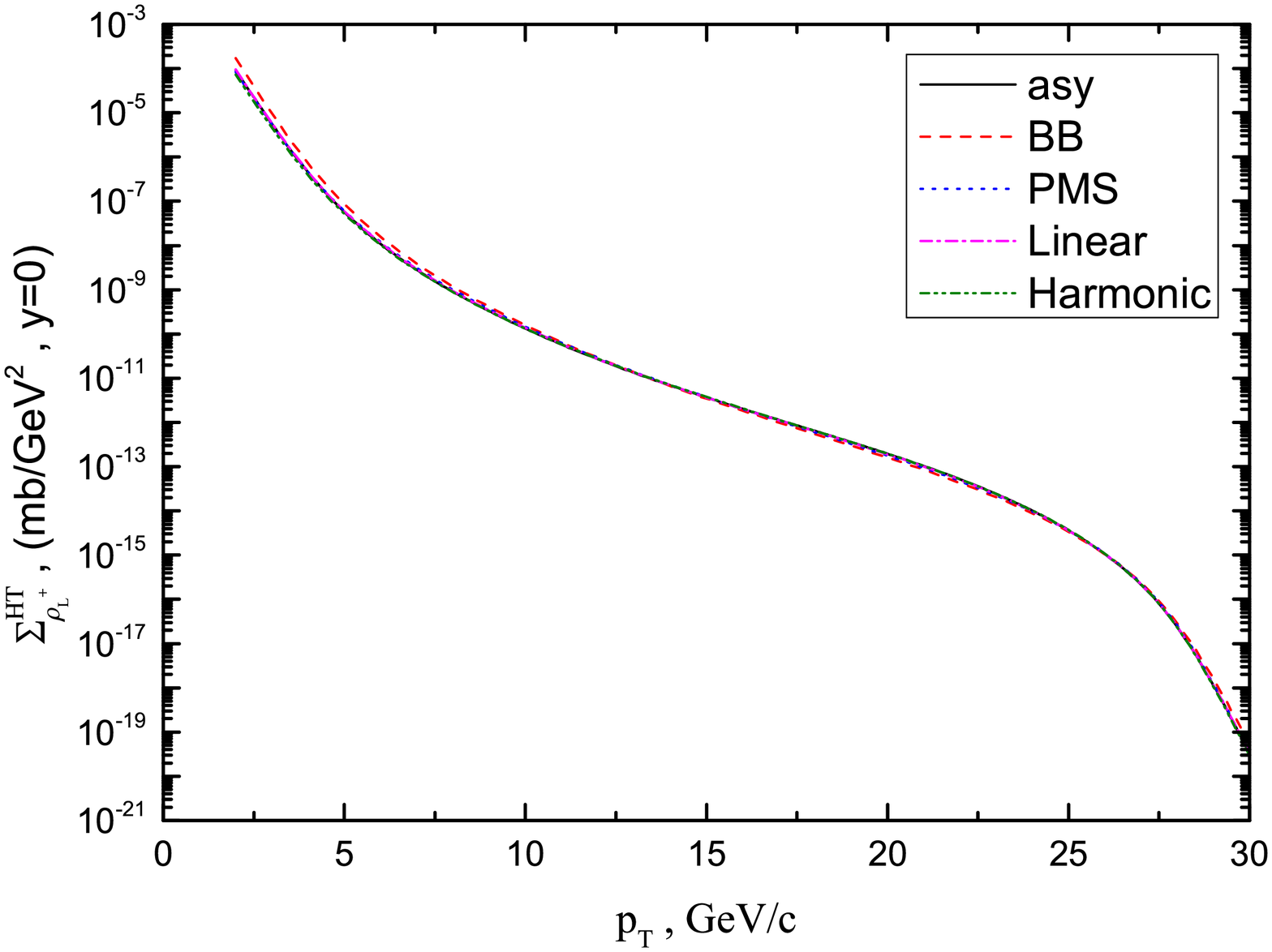}
\caption{HT contribution to $\rho_{L}^+$ meson production $p\bar
{p}\to\rho_{L}^{+}\gamma X$ cross-section $\Sigma_{\rho^+}^{HT}$ as
a function of the transverse momentum $p_{T}$ of the $\rho_{L}^+$
meson, at $\sqrt s=62.4$\,\, GeV and $y=0$} \label{fig:fig2}
\end{figure}
\begin{figure}[!htb]
\includegraphics[width=11.8 cm]{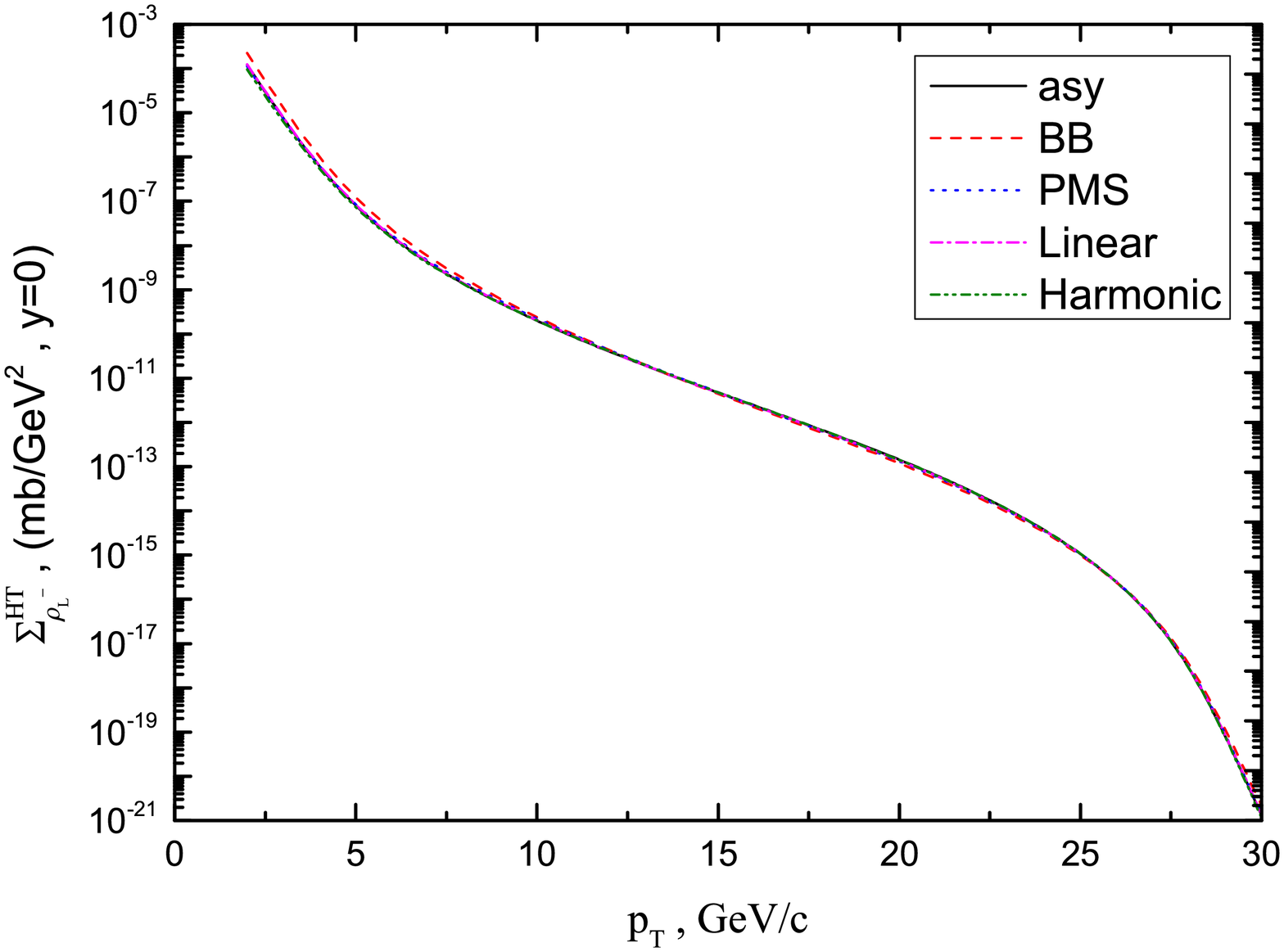}
\caption{HT contribution to $\rho_{L}^-$ meson production $p\bar
{p}\to\rho_{L}^{-}\gamma X$  cross-section $\Sigma_{\rho^-}^{HT}$ as
a function of the transverse momentum $p_{T}$ of the $\rho_{L}^-$
meson, at $\sqrt s=62.4$\,\, GeV and $y=0$} \label{fig:fig3}
\end{figure}
\begin{figure}[!htb]
\includegraphics[width=11.8 cm]{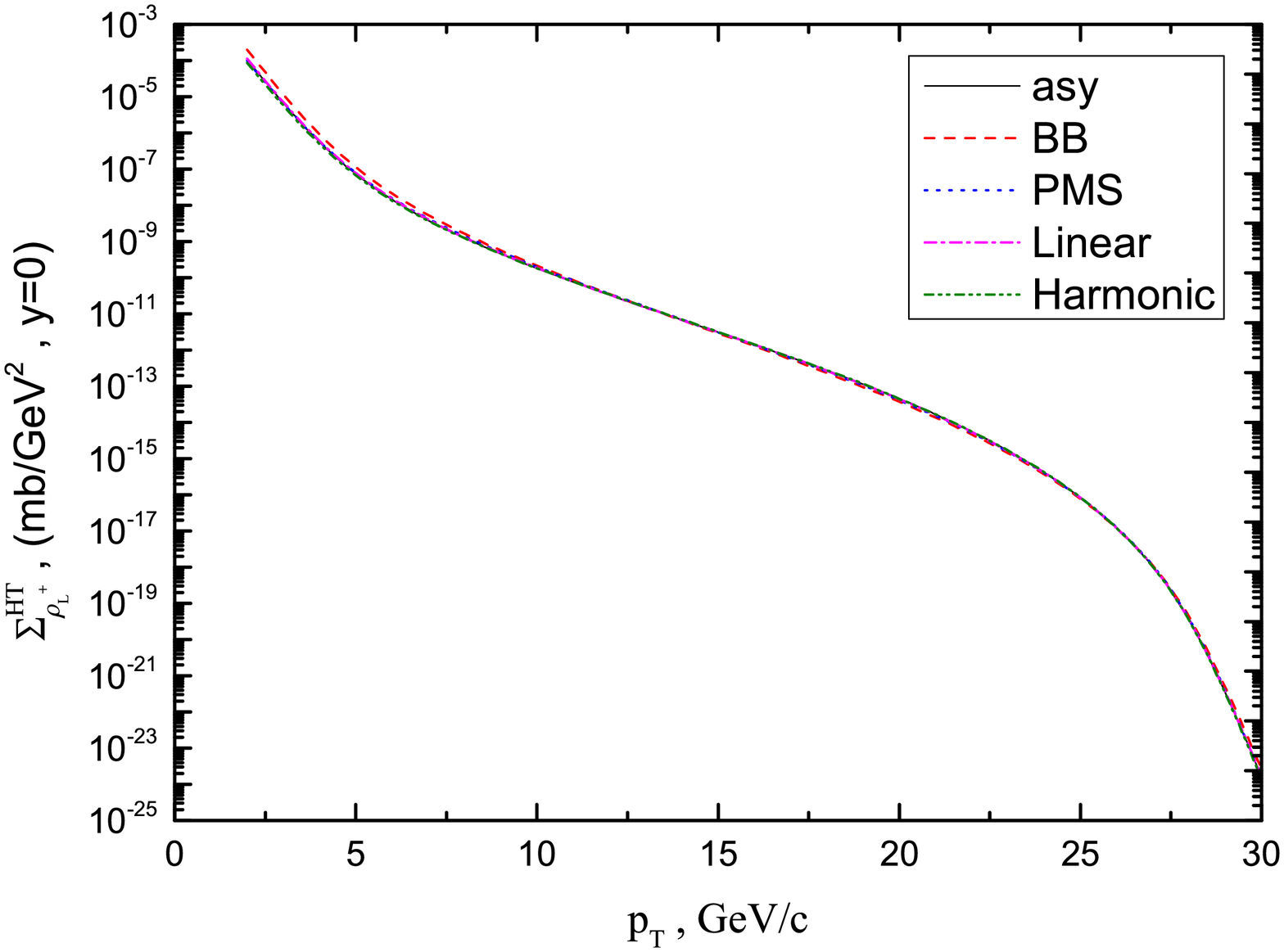}
\caption{HT contribution to $\rho_{L}^+$ meson production
$pp\to\rho_{L}^{+}\gamma X$  cross-section $\Sigma_{\rho^+}^{HT}$ as
a function of the transverse momentum $p_{T}$ of the $\rho_{L}^+$
meson, at $\sqrt s=62.4$\,\, GeV and $y=0$} \label{fig:fig4}
\end{figure}
\begin{figure}[!htb]
\includegraphics[width=11.8 cm]{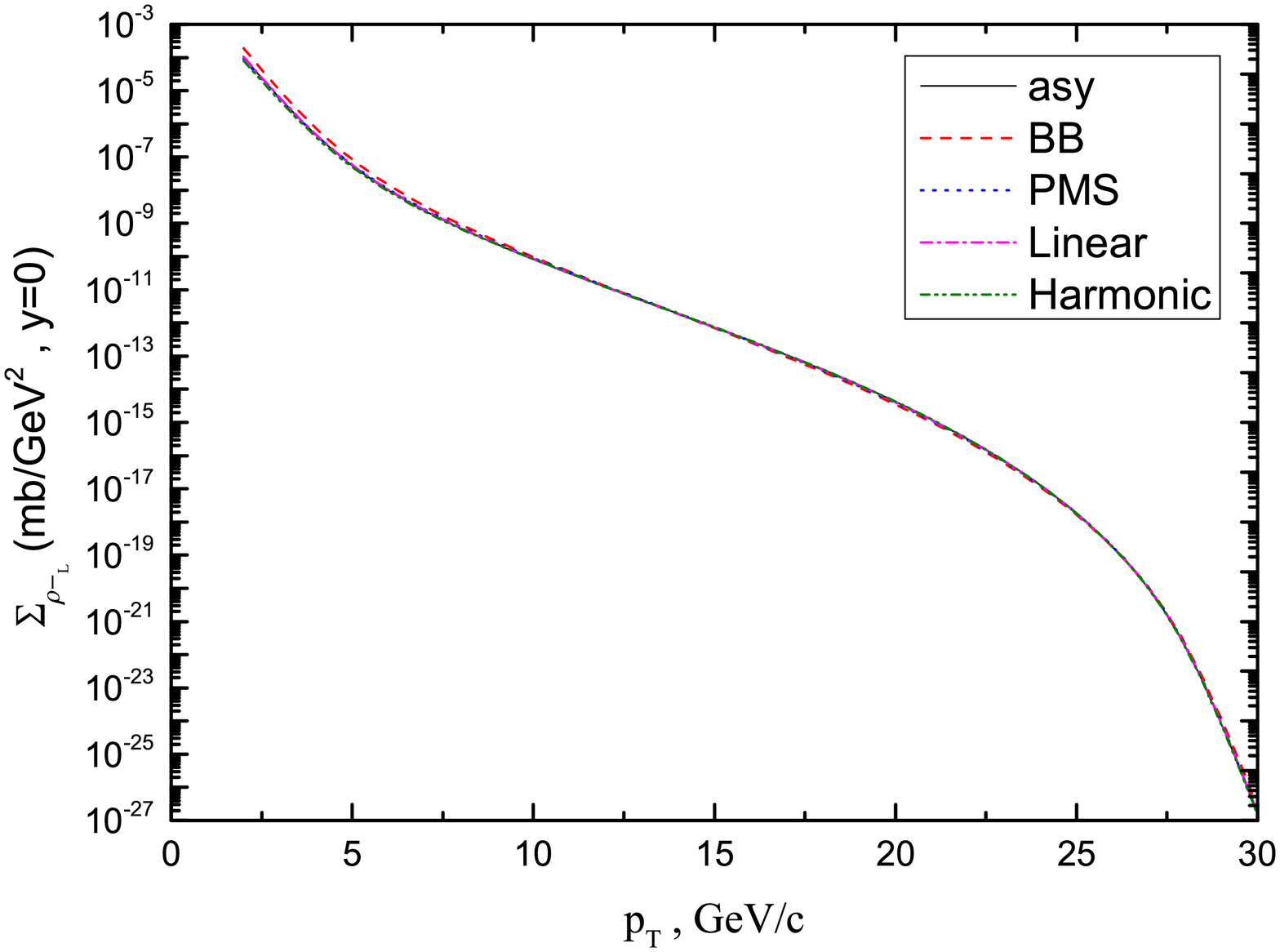}
\caption{HT contribution to $\rho_{L}^-$ meson production
$pp\to\rho_{L}^{- }\gamma X$  cross-section $\Sigma_{\rho^-}^{HT}$
as a function of the transverse momentum $p_{T}$ of the $\rho_{L}^-$
meson, at $\sqrt s=62.4$\,\, GeV and $y=0$} \label{fig:fig5}
\end{figure}
\begin{figure}[!htb]
\includegraphics[width=11.8 cm]{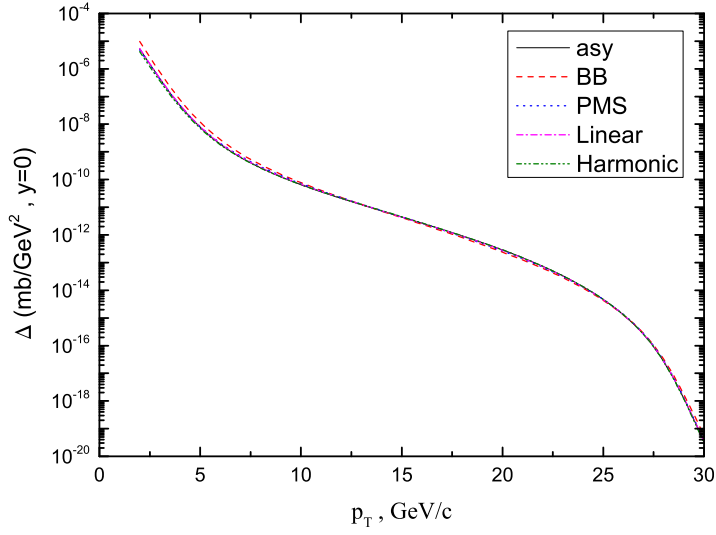}
\caption{HT contribution to difference of cross-section
$\Delta=\sum\nolimits_{p\bar p}-\sum\nolimits_{pp}$ as a function of
the transverse momentum $p_{T}$ of the $\rho_{L}^+$ meson, at $\sqrt
s=62.4$\,\, GeV and $y=0$} \label{fig:fig6}
\end{figure}
\begin{figure}[!htb]
\includegraphics[width=11.8 cm]{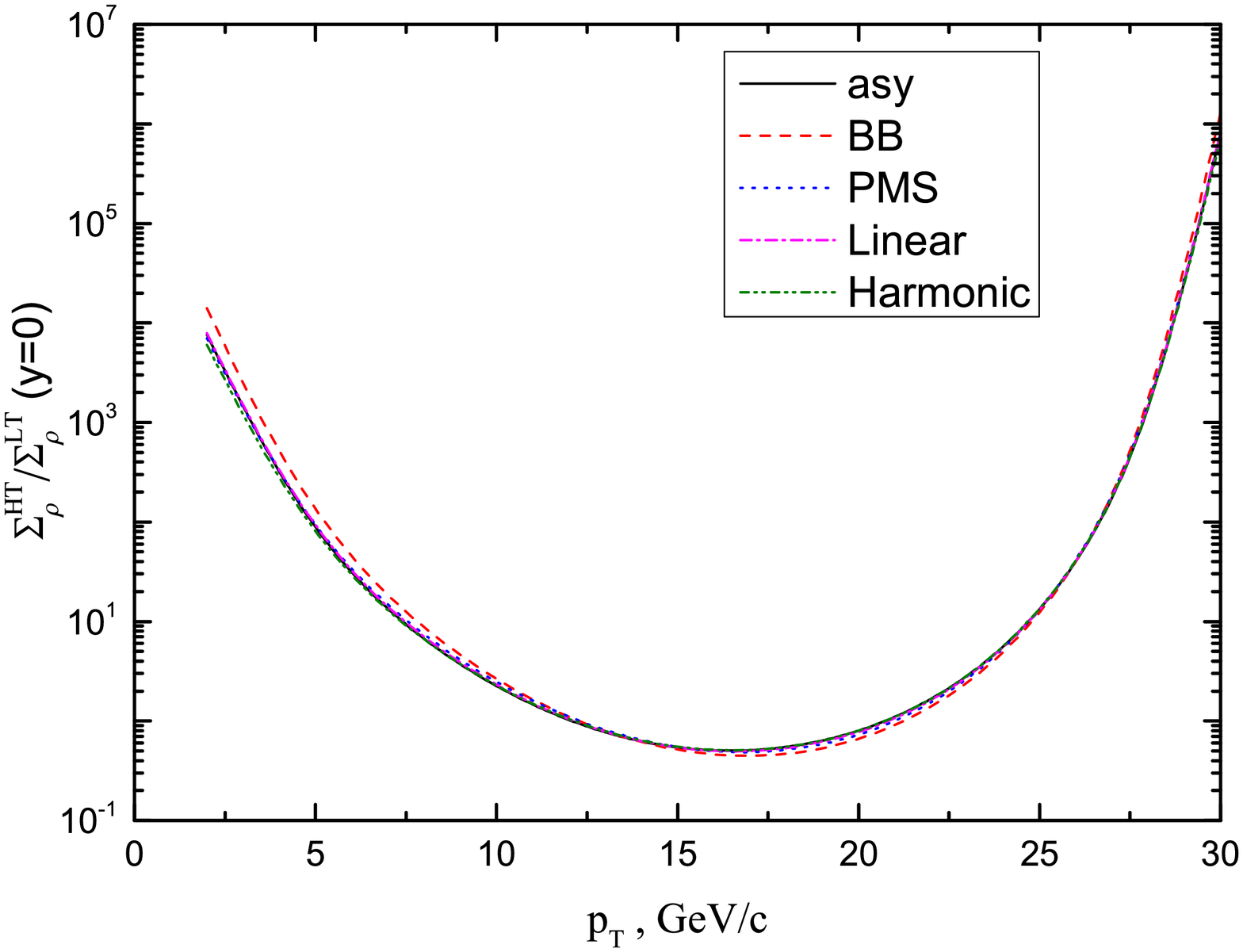}
\caption{Ratio $\Sigma_{p\bar p}/\Sigma_{\rho}^{LT}$ as a function
of the transverse momentum $p_{T}$ of the $\rho$ meson  at the c.m.
energy $\sqrt s$=62.4 GeV and $y=0$.} \label{fig:fig7}
\end{figure}
\begin{figure}[!htb]
\includegraphics[width=11.8 cm]{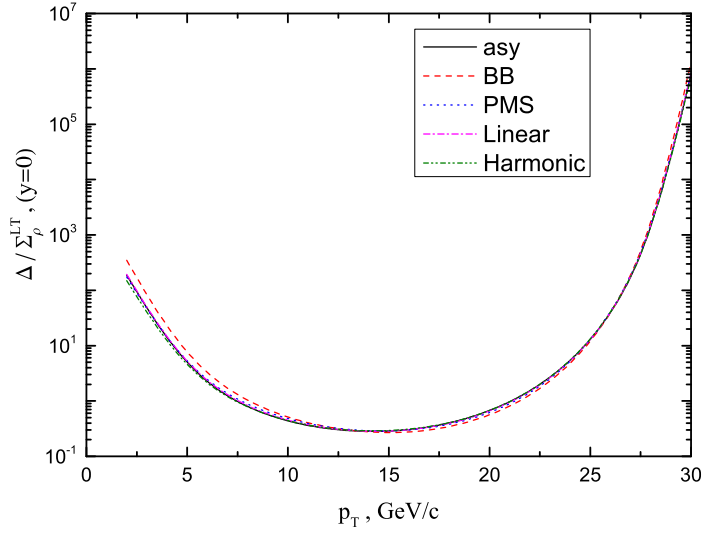}
\caption{Ratio $\Delta /\Sigma_{\rho}^{LT}$ as a function of the
transverse momentum $p_{T}$ of the $\rho$ meson at the c.m. energy
$\sqrt s$=62.4 GeV and $y=0$.} \label{fig:fig8}
\end{figure}
\begin{figure}[!htb]
\vskip-1.2cm \epsfxsize 11.8cm \centerline{\epsfbox{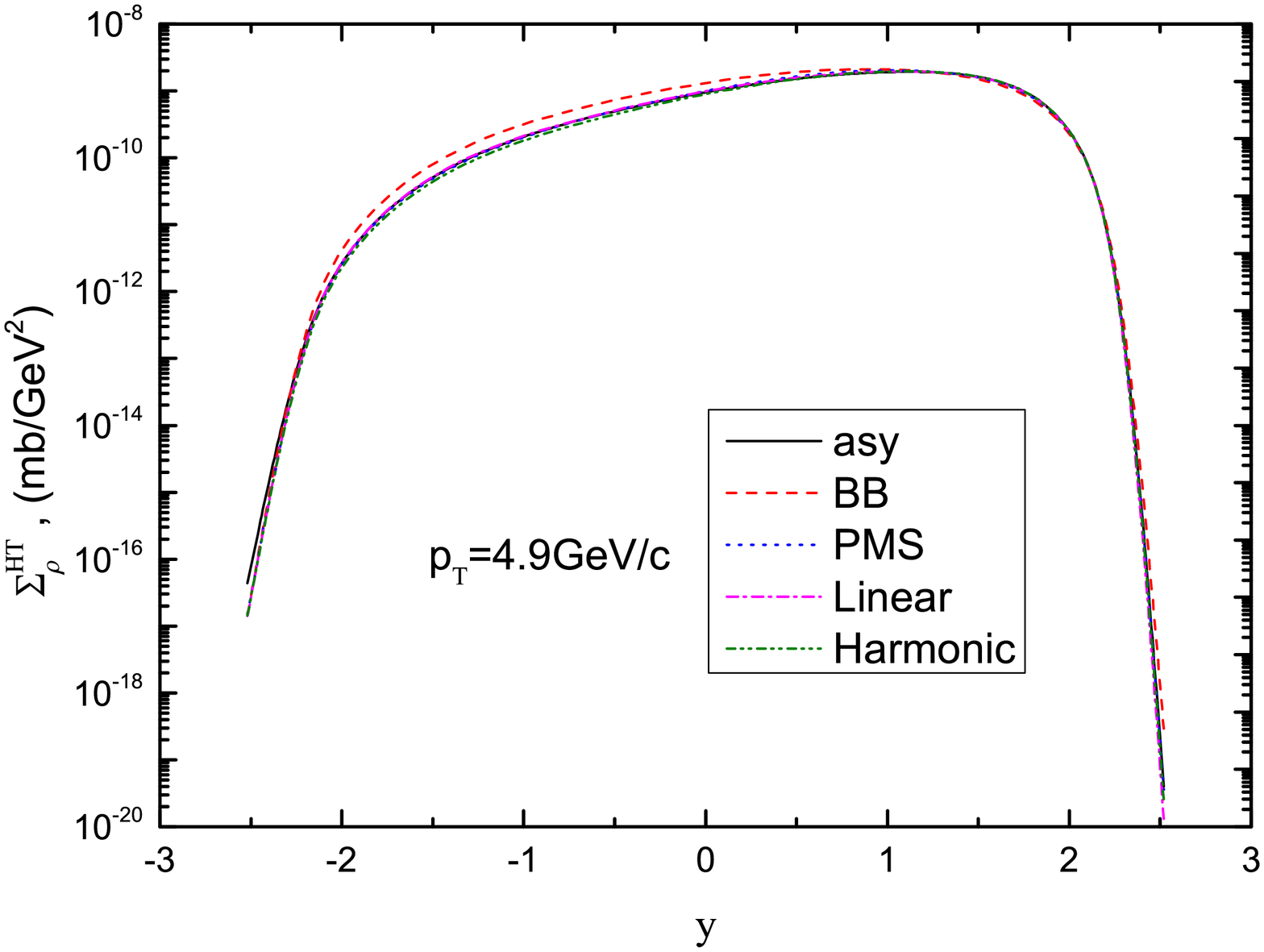}}
\vskip-0.2cm \caption{ HT contribution to sum of $\rho_{L}$ meson
production $p\bar {p}\to\rho_{L}\gamma X$  cross-section
$\Sigma_{p\bar p}$ as a function of the $y$ rapidity of the meson at
$p_T$=4.9\,\, GeV/$c$, at $\sqrt s=62.4$\,\, GeV.} \label{fig:fig9}
\end{figure}
\begin{figure}[!hbt]
\vskip-1.2cm \epsfxsize 11.8cm \centerline{\epsfbox{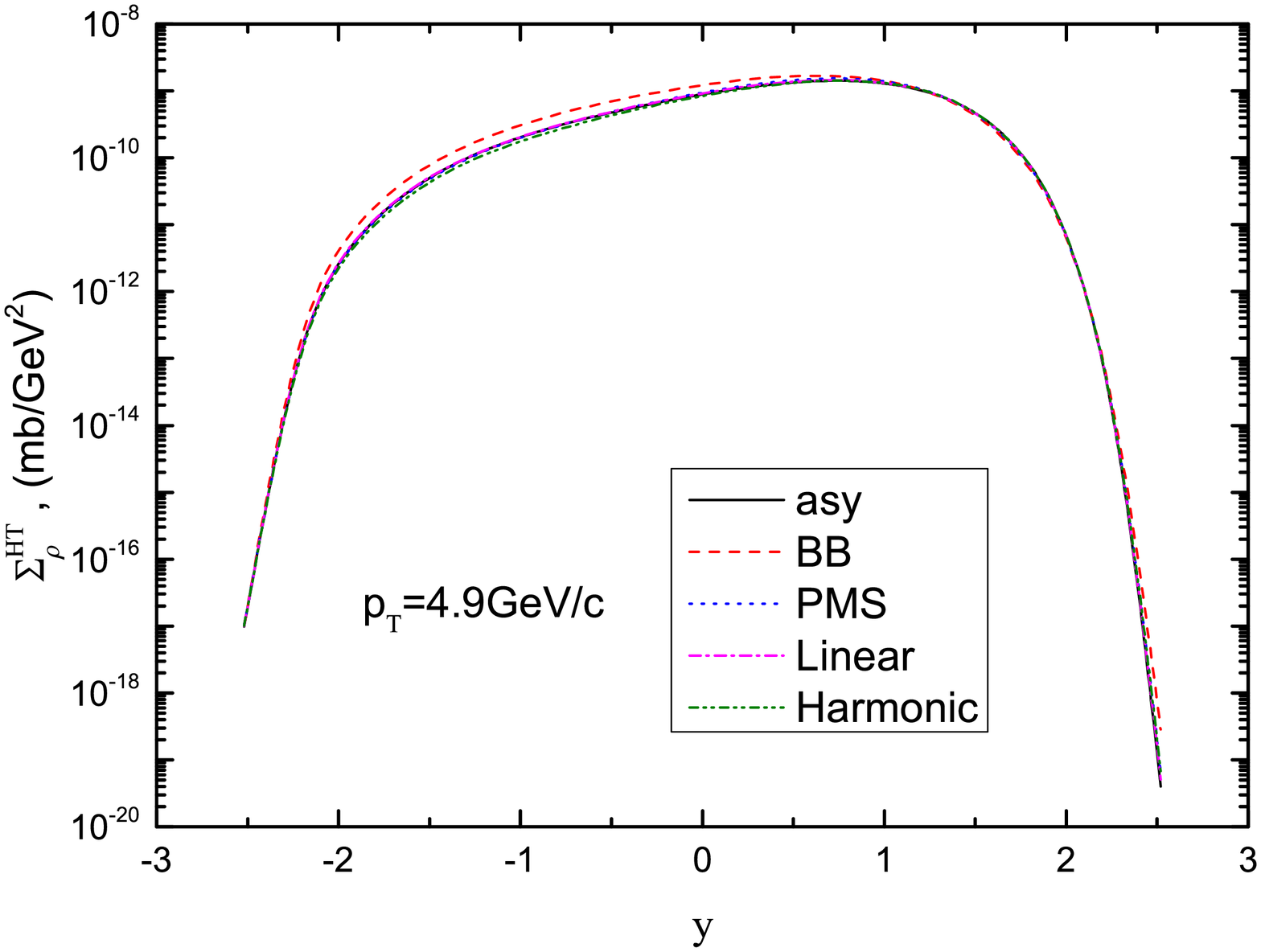}}
\vskip-0.2cm \caption{HT contribution to sum of $\rho_{L}$ meson
production $pp\to\rho_{L}\gamma X$  cross-section $\Sigma_{pp}$ as a
function of the $y$ rapidity of the meson at $p_T$=4.9\,\, GeV/$c$, at
$\sqrt s$=62.4\,\, GeV.} \label{fig:fig10}
\end{figure}
\begin{figure}[!hbt]
\vskip-1.2cm \epsfxsize 11.8cm \centerline{\epsfbox{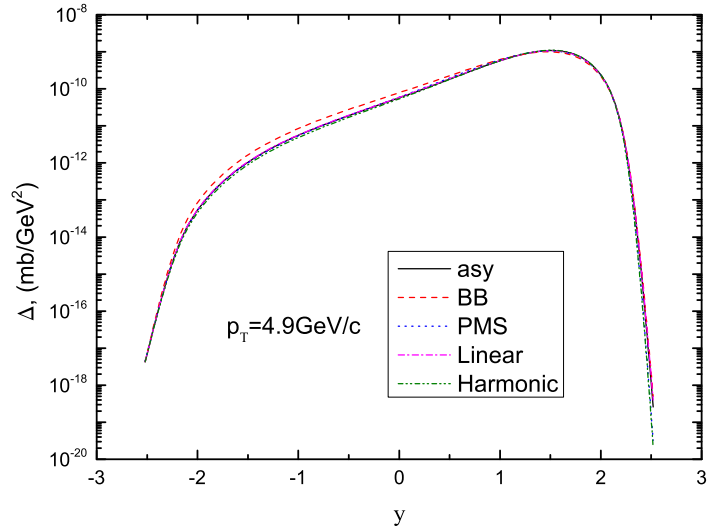}}
\vskip-0.2cm \caption{HT contribution to difference of cross-section
$\Delta=\sum\nolimits_{p\bar p}-\sum\nolimits_{pp}$ as a function of
the $y$ rapidity of the meson at $p_T$=4.9\,\, GeV/$c$, at $\sqrt
s$=62.4\,\, GeV.} \label{fig:fig11}
\end{figure}
\end{document}